\documentclass[%
reprint,
superscriptaddress,
 amsmath,amssymb,
 aps,
]{revtex4-2}

\usepackage{graphicx}% Include figure files
\usepackage{bm}% bold math
\usepackage{color} 
\usepackage{hyperref}

\usepackage{fancyhdr} 
\begin{document}

\title{Deep convection-driven vortex formation on Jupiter and Saturn}

\author{Rakesh Kumar Yadav}
\affiliation{Department of Earth and Planetary Sciences, Harvard University,  Cambridge, MA 02138, USA}
 \email{To whom correspondence should be addressed: rakesh\_yadav@fas.harvard.edu}

\author{Moritz Heimpel}%
\affiliation{Department of Physics, University of Alberta, T6G 2J1 Edmonton, Canada}

\author{Jeremy Bloxham}
\affiliation{Department of Earth and Planetary Sciences, Harvard University,  Cambridge, MA 02138, USA}

\date{\today}% It is always \today, today,
             %  but any date may be explicitly specified

\begin{abstract}
The surface of both Jupiter and Saturn has magnificent vortical storms which help shape the dynamic nature of their atmospheres. Land- and space-based observational campaigns over time have established several properties of these vortices, with some being similar between the two planets, while others are different. Shallow-water hydrodynamics, where the vortices are treated as shallow weather-layer phenomenon, is commonly evoked for explaining their formation and properties. Here we report novel formation mechanisms for vortices where the primary driving mechanism is the deep planetary convection occurring in these planets. Using three dimensional simulations of turbulent convection in rotating spherical shells, we propose two ideas: (1) rotating turbulent convection generates deep axially-aligned cyclones and anticyclones; (2) a deep planetary dynamo acts to promote additional anticyclones, some as large as Jupiter's Great Red Spot, in an overlying atmospheric layer. We use these ideas to interpret several observational properties of vortices on Jupiter and Saturn.\\
\\
\bf{Cite as: {\em Science Advances}, Vol.~6, no.~46, eabb9298 (2020); DOI: \href{https://doi.org/10.1126/sciadv.abb9298}{\tt 10.1126/sciadv.abb9298} }
\end{abstract}

\maketitle

\section{Introduction}

By modelling the dominant dynamical features present on the surface of Jupiter and Saturn, namely zonal jet streams and storms/vortices, we can learn about what drives them and their connection to the planetary deep interior. Apart from the well known and persistent hexagonal storm on Saturn and the Great Red Spot (GRS) on Jupiter, there are numerous other compact vortical storms with various sizes and life times found at various latitudes on these planets. By analyzing the images taken by the Cassini spacecraft through its lifetime, Trammel {\em et al.}~\cite{trammell2014, trammell2016} provide a detailed outlook on the properties of compact vortices on Jupiter and Saturn. In our study we focus only on the fluid dynamical properties of the vortices where the following general features stand out. First, with the exception of the equatorial regions containing a broad eastward zonal jet, vortices are generally found at all latitudes, but tend occur preferentially in regions of westward zonal flow \cite{vasavada2006, trammell2014}. There is a stark contrast in the number of vortices between the two planets: about 200 with 1000 km or larger diameter on Jupiter while only 10 to 50 on Saturn \cite{trammell2014}. Note that while both planets also have smaller vortices, the disparity in the numbers still persists \cite{li2004, vasavada2006}. Vortices are also present at or very close to the rotational poles: Saturn has a cyclonic polar vortex at each pole \cite{sanchezlavega2006} while Jupiter has a cluster of cyclones at each pole \cite{adriani2018}.

The rotation sense of vortices is also an important property but much harder to infer through direct observations. Li {\em et al.}~(2004) \cite{li2004} analyzed 500 vortices with diameter of at least 700 km on Jupiter using Cassini images. They inferred the vorticity of the largest 100 vortices, all of them being anticyclonic. An earlier study \cite{mac1986} using  Voyager images reached a similar conclusion. Due to the more hazy atmosphere of Saturn, the contrast of the different fluid dynamic features is much lower  compared with Jupiter. Paired with the lack of continuous images of Saturn's surface, determining the vorticity is not possible for most of the storms. Instead, the vorticity of a compact vortex is assumed to be determined by the vorticity of the local zonal wind shear \cite{vasavada2006, trammell2014, trammell2016}. Based on this assumption, Saturn's atmosphere does not appear to have a strong bias towards anticyclones. We must, however, keep in mind that the number of vortices on Saturn is small and undergoes large variations over time. For example, during the 7 year period from 2008 to 2015, the number of vortices of at least 1000 km diameter changed from about 5 to 20 in the northern hemisphere of Saturn \cite{trammell2016}. Therefore, a robust statistical trend cannot be inferred for Saturnian vortices.

\begin{figure*}
\includegraphics[scale=1.8]{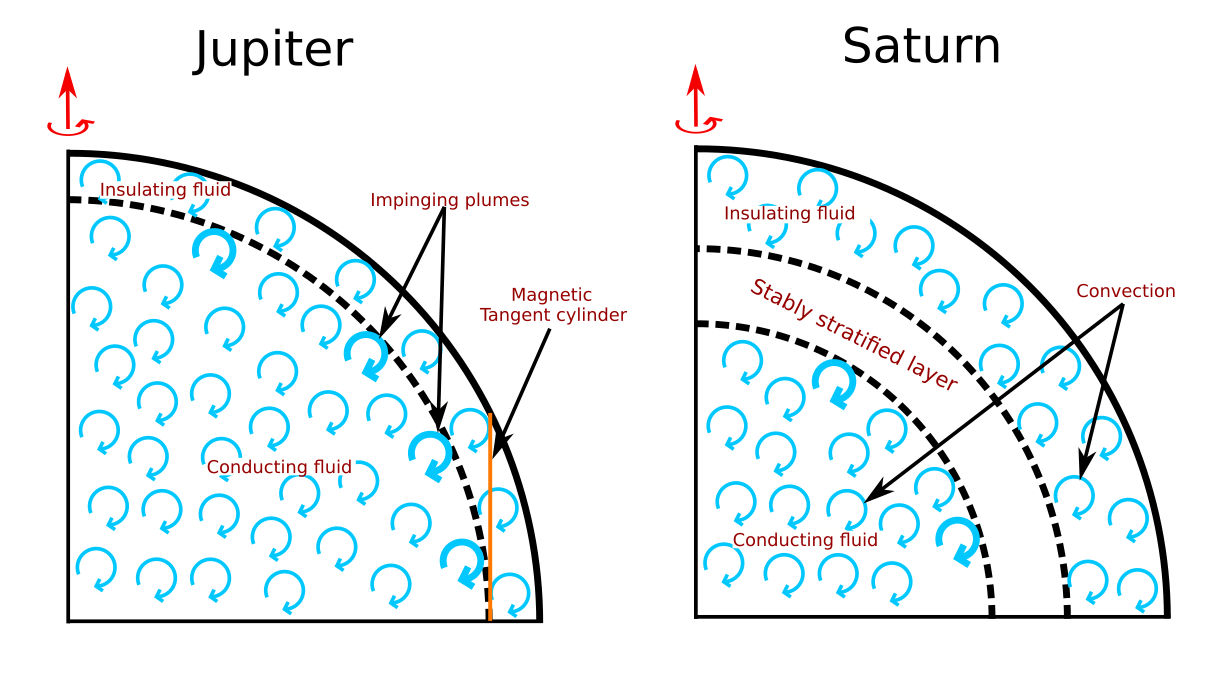}
\caption{\label{fig1} {\bf Gas giant interiors}:  A schematic representation of the interior of Jupiter and Saturn showing a meridional quadrant in the northern hemisphere. The partial circles with arrows represent overturning convection. The overturns with thick lines highlight energetic convection plumes that impinge on the adjacent layer. Jupiter may be crudely approximated by a two layer structure with the deep interior consisting of metallic hydrogen with larger conductivity, and a lower conductivity molecular hydrogen outer layer \cite{french2012}. Saturn, on the hand, likely has a three layer structure: an innermost metallic hydrogen layer sustaining the dynamo, a molecular hydrogen rich low conductivity outermost layer, and a  stably stratified layer due to Helium rain-out \cite{stevenson1982} between the other two layers. Although the exact location and nature of the intermediate Helium-enriched layer is uncertain, its existence is highly favored \cite{stevenson1982,stanley2010} due to the very high axisymmetry of Saturn's magnetic field \cite{cao2019}.  Jupiter may also have a similar stable layer but it is likely much thinner \cite{militzer2016}; furthermore, Jupiter's magnetic field is not highly axisymmetric \cite{connerney1982,connerney2018}. }
\end{figure*}

\begin{figure*}
\includegraphics[scale=0.57]{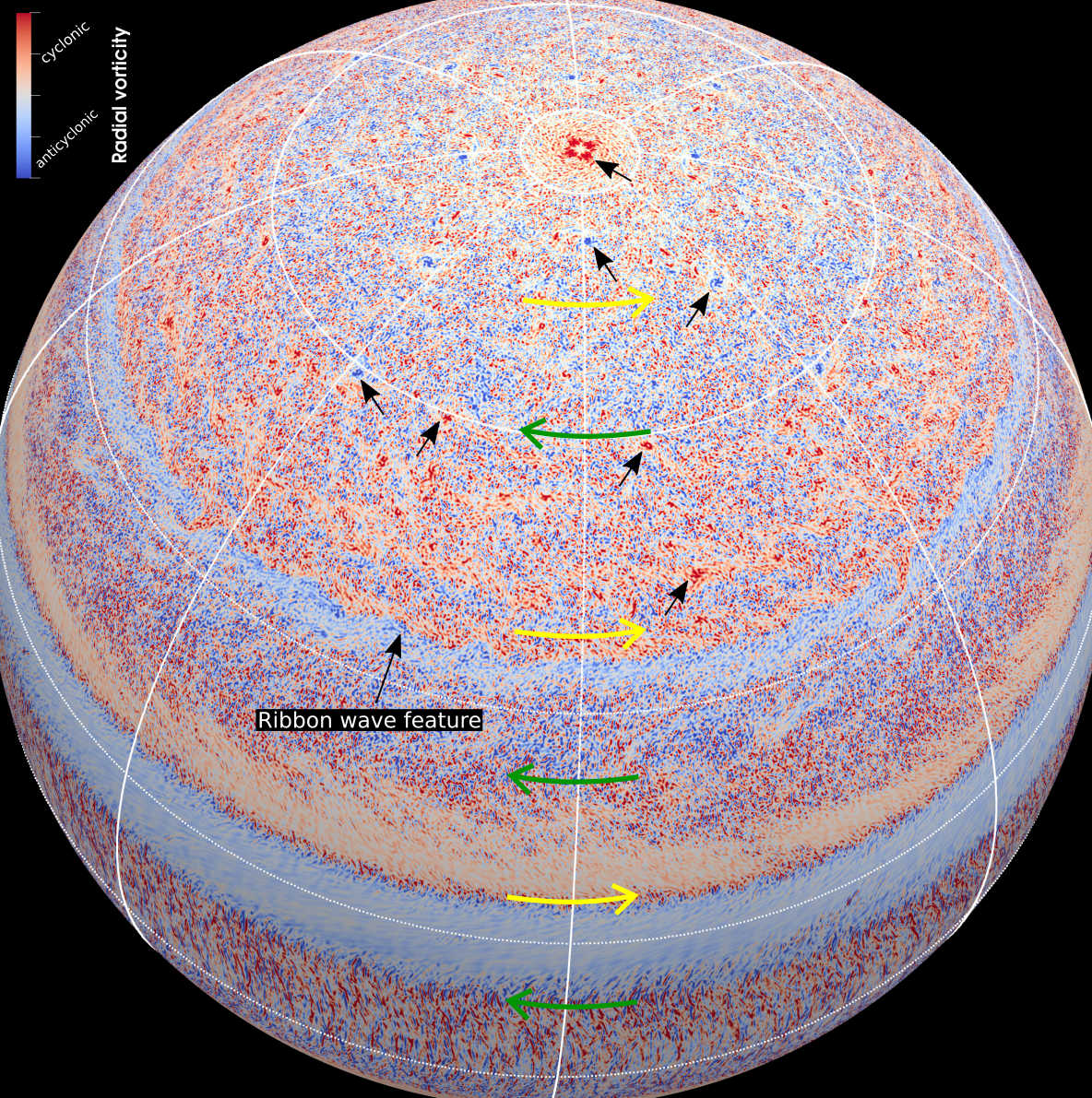}
\caption{\label{fig2} {\bf Cyclones and anticyclones}: The radial component of the vorticity, defined by $(\nabla\times\bf{u})_r$, where $u$ is velocity, plotted at mid depth of the simulation. The velocity is in the units of Rossby number $u/\Omega r_o$, where $\Omega$ is planetary rotation frequency, and $r_o$ is shell radius. The (-ve) +ve values are represented by (blue) red color and the plotted range is $\pm$0.006. The (yellow) green arrows highlight the peaks of (eastward) westward zonal jets. Some of the easily recognizable cyclonic (red) and anticyclonic (blue) vortices are pointed out with arrows, as well as a large scale modulation labelled `Ribbon wave feature'.  Note that when a quarter wedge is repeated to form a full spherical surface, as is done here, a vortex close to the poles would appear as a cluster of four vortices. This is most likely an artifact since in our other simulations where we considered full azimuthal extent only one cyclonic vortex was present at or near the poles \cite{yadav2020}.}
\end{figure*}

Much like in the study of zonal jets on gas giant planets (see \cite{cho1996b, williams2003, liu2010} for shallow jets and \cite{busse1976,christensen2001,aurnou2001,heimpel2005,gastine2014zon} for deep jets), vortices can also be understood as either existing only in the outermost weather layers or extending much deeper into the planetary interior. In the case of zonal jets, exciting progress was recently made by modelling the gravity data from the Juno mission and the Cassini Grand Finale. It indicates that the zonal jet streams on both planets are likely thousands of kilometers deep, about 9000 km in Saturn \cite{galanti2019} and about 3000 km in Jupiter \cite{kaspi2018}. These results lend support to the deep zonal jet scenario where planetary convection is the driving force. Given these results, it is certainly worth investigating whether some of the vortices visible on the surface extend deep into the interior. 

So far, most vortex formation models consider only the thin outer weather layer, with several using the shallow-water equations with a prescribed forcing in the form of moist convection or solar heating \cite{pedlosky1987, vasavada2005}. Such studies have successfully modelled several properties of the vortices, for example, the bias towards anticyclones  \cite{arai1994, polvani1994} and polar vortex formation \cite{oneill2015, brueshaber2019}. However, it is also well known that coherent vortices can spontaneously form in three dimensional fluid turbulence if rotational effects are significant (e.g., see \cite{alexakis2018}). Could  deep planetary convection also generate coherent vortices along with zonal jets? There have been several attempts to study this. Coherent, primarily anticyclonic, vortices have previously been shown to coexist with zonal jets using anelastic, spherical deep-convection models with a stably stratified layer \cite{heimpel2016}. Others have investigated vortex formation in Cartesian geometry \cite{chan2013, rubio2014, guervilly2014} where the curvature effects due to the spherical shell are not present. Here, we report several physically-motivated simulations of planetary convection in deep spherical shells to better understand how vortices form and behave under such conditions.

\section{Results}
\subsection{Thin shell case}

Keeping the general features of the giant planet interiors in mind (Figure 1), we explore two cases which will help us to better understand vortex formation in rotating spherical shells. We first consider the dynamics in a thin rotating spherical shell --  a generic representation of the outer convective layers in giant planets that couple only weakly with the interior magnetic field \cite{heimpel2005,jones2009} -- which spans a region from 0.97$r_o$ to $r_o$, where $r_o$ is the outermost radius of the shell. Such a shell thickness would be equivalent to about 2000 km and 1000 km if scaled to Jupiter and Saturn, respectively.  Since contemporary 3D simulations can not attain the highly turbulent geostrophic regime present in the interiors of giant planets, we need to set up the simulation such that it can promote relevant physics. The choice of assuming a relatively thin layer is motivated by the fact that 2D and quasi-2D rotating flows are well known for vortex formation in a broad parameter regime \cite{boffetta2012}.  The thin shell nature of the simulation will allow it to excite vortices but still retain the 3D nature of the convective flows. We ignore the fluid compressibility in this simulation for the sake of simplicity and to reduce the computational costs associated with the model. Furthermore, a four-fold symmetry is imposed in the azimuthal direction to limit the computational requirements. The crucial control parameters governing the employed non-dimensional Boussinesq equations are: the Ekman number defining the ratio of viscous and Coriolis forces is $10^{-5}$, the Rayleigh number defining the convective driving is $3\times10^7$, and the thermal Prandtl number is 0.1. Additional model details can be found in the {\em Materials \& Methods} section.

This thin layer model self-consistently generates about seven alternating zonal jets in each hemisphere with an eastward jet in the equatorial region (Supplementary Figure S1), as well as a number of compact vortices. Alternating zonal jet streams have already been reported in earlier simulations of spherical shells representing the deep convection in the outer atmospheres of giant planets (e.g.~see \cite{christensen2001,aurnou2001,heimpel2005,gastine2014zon}). Furthermore, anticyclonic vortices, along with zonal flows, were also reported earlier \cite{heimpel2016} in a similar but thicker shell setup with an overlying stably stratified layer.

The simulation generates about 40 well defined, plus additional smaller and shorter-lived, vortices in each hemisphere (Figure 2). The vortices are formed in the high to mid-latitude regions (Supplementary Figure S2) and the lower latitudes are dominated by stronger zonal jets. This may be expected since the topographic $\beta$-effect (i.e., changing height of an axially aligned fluid column with latitude) is larger at low latitudes which promotes the development of zonal jets \cite{rhines1975}. Around mid latitude a clear wavy pattern is present in the vicinity of the peak of an eastward jet (Figure 2) that roughly separates the low-latitude jet regime from the high latitude vortex regime. This feature is reminiscent of the `Ribbon wave' observed on Saturn \cite{smith1982} at similar latitudes. A persistent cyclonic vortex was present throughout the simulation at or close to both rotational poles.

The dynamics present in the simulation is very rich and can be better appreciated by visualizing a time evolution of the flow; the Supplementary Animation 1
(\href{https://youtu.be/bmBdI9sJbqc}{\tt https://youtu.be/bmBdI9sJbqc}) shows the time evolution of mid-depth radial vorticity over 80 rotations. Well defined vortices can be easily tracked as they are advected by the background zonal flow. The vortex life time appears to be mainly dependent on the size, with smaller ones lasting for several days while some large ones, e.g.~the anticyclone highlighted by the second black arrow from the north pole in Figure 2, persisting for the entire duration of the animation. During this time evolution, this stable anticyclone experienced a slow drift to lower latitudes. The animation shows instances where storms with the same sign of vorticity merge with each other, leaving a single storm, as well as where storms with opposite vorticity sign merge and vanish. In some instances, storms come very close to each other but are advected away quickly by the background zonal flow. Generally, both cyclones and anticyclones form all over the mid-to-high latitudes, however, (anti)cyclones dominate in the region where the local zonal flow shear direction is (anti)cyclonic. This tendency is evident in the 65N to 85N and in the 50N to 65N latitude ranges where a preference for anti-cyclones and cyclones, respectively, is visible. Such a vortex dynamics is reminiscent of the interactions of storms on both Saturn and Jupiter \cite{vasavada2005}. 

We contrast the results from this thin shell case with another study by Heimpel {\em et al.} (2016) \cite{heimpel2016} that also reported vortices in a rotating spherical shell with convective forcing. There, all the vortices were anticyclonic due to the underlying formation mechanism: a stably stratified thin layer in the outermost part of the shell makes radially outward convective plumes to diverge, which, later, form shallow anti-cyclonic circulations in the outer layers due to the Coriolis force. This vortex formation mechanism is different from the one in our simulation where the vortices are self-organization phenomenon promoted by the inverse cascade of energy as seen in earlier Cartesian box studies \cite{chan2013, rubio2014, guervilly2014}. Storms in our simulation are deep geostrophic structures and change little along the direction of the planetary rotation (Supplementary Figure S3). Furthermore, the shallow stably-stratified-layer mechanism \cite{heimpel2016} only produces anti-cyclones, whereas, in our simulation both cyclones and anticyclones form depending on the local zonal flow shear. 

Comparing this case with other simulations we performed with thicker spherical shells (not reported here) leads us to believe that the thin shell nature of the shell is helping to promote the self-organization of compact vortices. As mentioned earlier, the main idea behind choosing a thin layer is to promote vortex formation in the parameter regime we can access. We know that 2D layers can promote vortices in a broad parameter regime \cite{boffetta2012} but lack the complex convection dynamics. In a deeper spherical shell simulation setup (akin to Saturn’s deep atmospheric layer) we can not achieve the highly turbulent and geostrophic regime that is likely needed to excite self-organization of vortices. Therefore, our thin shell setup, retaining properties of both 2D and 3D setups, is a good compromise to simulate relevant physics.

\subsection{Thick shell case}

\begin{figure*}
\includegraphics[scale=0.46]{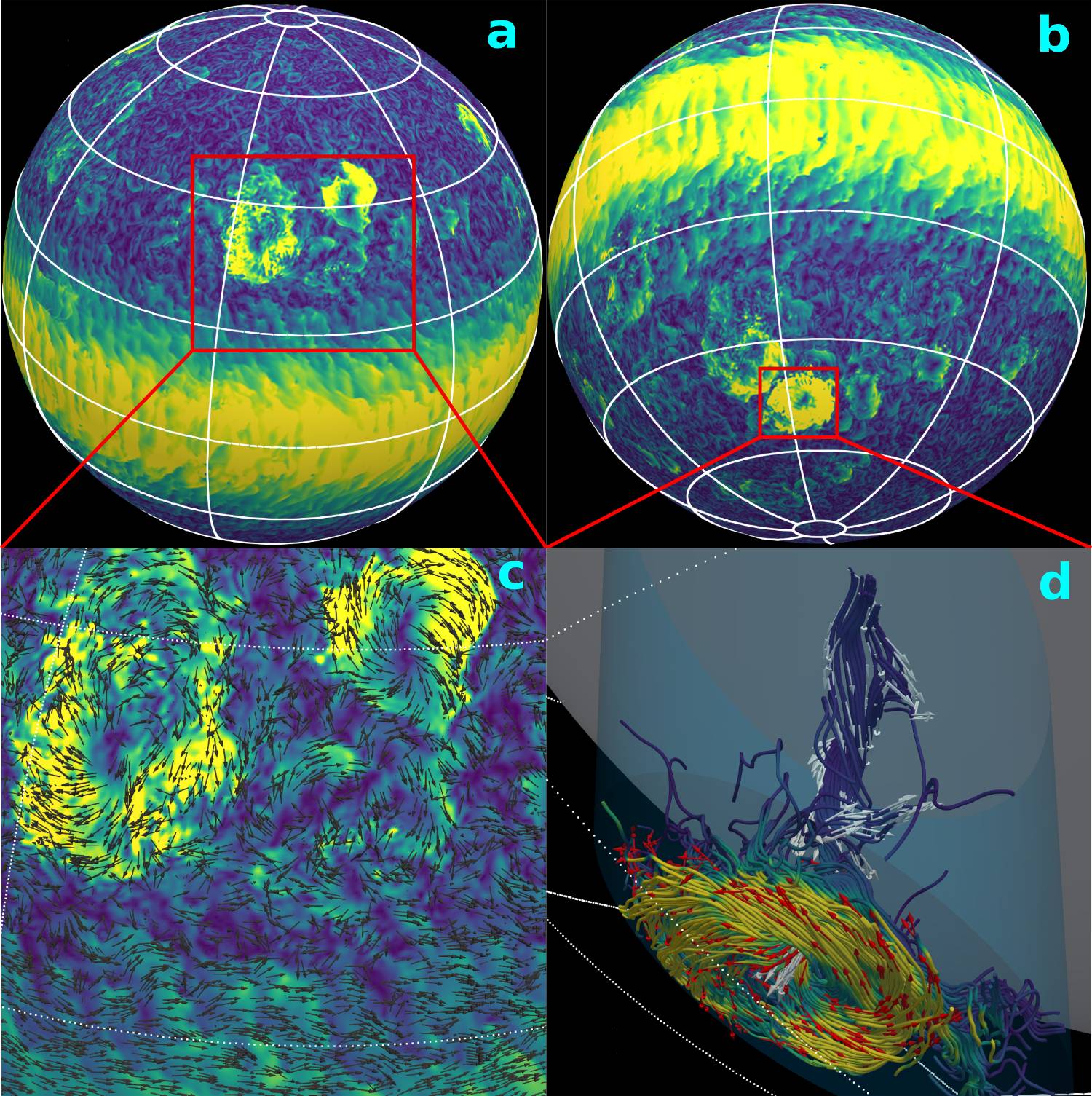}
\caption{\label{fig3} {\bf Deep rooted anticyclones}: The velocity magnitude on the outer boundary of the simulation from a northern (panel {\bf a}) and southern (panel {\bf b}) perspective. Lighter colors represent larger velocities with maximum being 0.05 in the units of Rossby number. Panel {\bf c} shows the velocity magnitude as well as flow vectors in a subset of the region highlighted with a box in panel {\bf a}. Panel {\bf d} shows the flow streamlines and flow vectors -- white ones for the convective plume triggering the storm and red ones for the rim of the storm -- for an anticyclone highlighted with a box in panel {\bf b}. The dark gray surface in panel {\bf d} highlights the transition radius where the electrical conductivity starts to drop. Furthermore, the translucent cylinder in panel {\bf d} is aligned with the rotation axis.}
\end{figure*}

Next, we investigate what happens when a hydrodynamic layer directly interacts with another layer where the electrical conductivity is large enough to allow generation of magnetic fields (Figure 1). We consider a rotating spherical shell that spans a region from 0.1$r_o$ to $r_o$, within which the fluid density decreases by a factor of about 150 along the radius in the shell. Similar to Dietrich \& Jones (2018) \cite{dietrich2018}, the electrical conductivity of the fluid changes with radius as a hyperbolic tangent function: it is constant until $\approx 0.84r_o$, then sharply decreases by seven orders of magnitude within $\approx 0.07r_o$, and stays constant afterwards (see `Profile 1' in Supplementary Figure S4). Such an electrical conductivity profile defines a deep dynamo layer and an overlaying atmospheric layer where Lorentz forces will be negligible. Therefore, we label the radial level at $0.85r_o$ as  the `transition' radius separating the dynamo and the atmospheric region. Note that our chosen electrical conductivity profile differs from several earlier studies: the hyperbolic tangent function introduces sharper conductivity gradients in the transition radius as compared to a combination of constant for some radius and exponentially decreasing afterwards \cite{gomez2010,duarte2013,jones2014,gastine2014dyn}. The electrical conductivity in Jupiter starts to decay with a super-exponential rate after about 0.9$r_o$ \cite{french2012} which leads to decoupling of the atmosphere from the dynamo in a relatively small radius range. In order to capture such a decoupling, it is crucial to include a strong electrical conductivity decay offered by a hyperbolic tangent function with a large step change. We then solve the convective magnetohydrodynamics in this setup. The crucial parameters governing the anelastic system of equations are: the Ekman number is $10^{-6}$, the Rayleigh number is $4\times10^9$, the thermal Prandtl number is 0.1, and the magnetic Prandtl number is 1.5. We refer the reader to the {\em Materials \& Methods} section for more details about the model. 

In this two-layer setup, the dynamics and the emergent features are very different from the earlier thin shell case. The simulation generates a largely dipole dominant magnetic field in the interior dynamo region (Supplementary Figure S5). Unlike the previous simulation exhibiting multiple alternating zonal jets, here we only get one equatorial eastward jet extending from about 30 degree north to 30 degree south (Supplementary Figure S6a). Investigating the zonal flow profile throughout the interior helps to explain the surface behavior. The strong Lorentz forces associated with the deeper dynamo layer enforce a nearly rigid body rotation in the fluid below $\approx0.85r_o$. Above this radial level, a hypothetical {\em magnetic} tangent cylinder (MTC; Figure 1), that is tangent to the transition radius and parallel to the rotation axis, demarcates two regions \cite{dietrich2018}. For cylindrical radii larger than MTC,  i.e.,~outside the MTC, the flow sharply decouples from the magnetic field due to much lower conductivity and creates a strong eastward zonal jet. For smaller cylindrical radii, i.e., within the MTC, the zonal flow is weak despite the flow-field decoupling. This property can be understood as following: an axially aligned fluid column in region outside the MTC experiences very little friction from the free slip mechanical boundaries imposed on its two ends, allowing it to develop strong zonal flows; inside the MTC, however, although the fluid column has a free-slip mechanical boundary on the upper end, it has a dynamo-imposed nearly rigid rotation on the lower end that mimics strong boundary friction, prohibiting the development of zonal flows \cite{aurnou2001,duarte2013}. We note that Jupiter and Saturn have alternating zonal flows at all latitudes. This is not a contradiction. It is generally believed that, unlike in the contemporary high-viscosity simulations, the zonal jets on Jupiter and Saturn experience much smaller drag due to the extremely small fluid viscosity \cite{french2012}. This allows the jets to readily develop at all latitudes in the molecular envelope of these planets despite having a dynamo-imposed nearly rigid body rotation in the deep \cite{guillot2018}. Recent laboratory experiments on jet formation with low viscosity fluids lend support to these ideas \cite{cabanes2017}.

Analyzing the flow pattern on the outer boundary of the simulation reveals interesting dynamics. The most dominant feature on the surface is the eastward jet in the equatorial region. In mid-to-high latitudes energetic vortical structures are present, which stand out in the otherwise uniformly distributed background of convective plumes (Figure 3). The energetic vortices in both the north and the south latitudes are anticyclonic in nature. A closer inspection of the vortices reveals that they have a quiet core with small velocities, surrounded by a large fast-rotating rim (Figure 3c, top right). Some of the large vortices also show a more complex behavior where a clear center and a rim are less well defined, but overall the structure constitutes an anticyclonic vortex (Figure 3c, top left).  These vortices form on scales ranging from 5 to 20 degrees in latitudinal span (equivalent to about 6000 to 24000 km in diameter if present on Jupiter's surface). The life time also varies with small storms persisting for only one planetary rotation while some larger ones lasting for 20 or so (see Supplementary Animation 2; \href{https://youtu.be/Azlfva56jb8}{\tt https://youtu.be/Azlfva56jb8}). Some of the long lasting storms at low latitudes also experience a westward drift of about 0.5 deg/day. 

Investigating the structure of a vortex in deeper layers reveals the underlying formation mechanism: as a convective plume rises from deeper depths it impinges on the outer boundary and diverges, the background Coriolis force then transforms the horizontal divergence to an anticyclonic circulation (Figure 3d). Therefore, although the anticyclones are relatively shallow and have a pancake shape, they have a very deep origin. 

Presence of such large anticyclonic vortices has not been reported in conventional spherical shell simulations of the outer convective layers of giant planets \cite{busse1976,christensen2001,aurnou2001,heimpel2005,gastine2014zon}, therefore, the deeper dynamo layer is the key for triggering them. The electrical conductivity jump at the transition radius essentially couples two dynamically distinct layers.  The convection in the deep dynamo continuously launches plumes into the overlying layer. Some of these plumes manage to either cross the entirely of the atmospheric layer and impinge on the outer layer or interact with the upper layer to create plumes that go on to hit the outer boundary and create anticyclones. Since the plumes are coming from deeper denser regions, when they deposit their energy in the outer less-dense layers, they can create substantially stronger flows than the surrounding local flows. The importance of the dynamo layer is further underscored by the fact that some of the low-latitude long lasting storms experience a slow westward drift which is driven by the weak westward zonal flow produced in the dynamo region that permeates the atmospheric layer (Supplementary Figure S6b). Such westward drifts are commonly found in dipole-dominant dynamos \cite{aubert2005,schaeffer2017}. Furthermore, the sharp conductivity gradient in our model is also an important ingredient to excite the anticyclones, another simulation with a more conventional profile of constant conductivity till $0.85r_o$ and an exponential drop afterwards (`Profile 2' in Supplementary Figure S4) did not produce similar anticyclones. This is likely due to the presence of a gradual transition of the fluid dynamics from the dynamo layer to the hydrodynamic layer which makes it one system rather two layers interacting with each other. Since the anticyclones are formed in the outer low-conductivity layer, they do not have a corresponding magnetic signal associated with them (Supplementary Figure S7). However, the simulation does have a tendency to generate more anticyclones in the regions containing higher magnetic energy. This hints that magnetic Lorentz forces are likely playing a role in promoting the formation of the anticylones.

\section{Discussion}

We have presented two simulations of highly turbulent convection in rotating spherical shells. The first simulation demonstrates the spontaneous formation of zonal jets, cyclones, and anticyclones in a thin spherical shell. The second one shows a novel vortex formation mechanism where plumes originating in a deep dynamo layer exclusively excite anticyclones in an overlying low-conductivity atmospheric layer. 

The results from these simulations can help us to better understand the atmospheric dynamics visible on the surface of Jupiter and Saturn.  Although vortex formation (through an inverse cascade of energy) is well known in rotating Cartesian boxes with either forced or convective turbulence \cite{chan2013, rubio2014, guervilly2014}, our simulation is the first  -- barring Heimpel {\em et al.}~(2016) \cite{heimpel2016} who report anticyclones in a stably stratified layer above a convective layer -- to produce both cyclones and anticyclones, as well as zonal jet streams, in a global spherical geometry with convective driving. Compared to  local Cartesian box simulations, our thin shell simulation can be considered a step further towards the goal of modelling the {\em global} atmospheric dynamics of gas giant planets. The global geometry allows us to simultaneously capture the vortex-free near equatorial regions, alternating zonal jets, formation of storms with vorticity governed by local shear flows, and non-linear interactions between storms (including mergers). The simulation shows that rotating convection can generate deep axially-aligned storms that mimic many properties of the storms on Jupiter and Saturn. 

The second case shows how a dynamo layer might interact with an overlaying atmospheric layer. The dynamo region is conventionally viewed as an inert layer proving a drag force for the atmospheric zonal jets \cite{christensen2001,aurnou2001,heimpel2005,gastine2014zon}. Our simulation, instead, shows that it could play an important role dynamically and may provide a source for seeding and sustaining large anticyclones in the atmospheric layer. Given that, there is a possible connection between our results and Jupiter's Great Red Spot (GRS). Jupiter's deep dynamo is expected to be in the magnetostrophic regime where small scale as well as planetary-scale giant convective cells are expected \cite{chandrasekhar1961,aurnou2017, yadav2016, aubert2017}. If Jupiter's dynamo layer does indeed seed anticylones in the atmospheric layer, as suggested by our simulation, then a GRS type anticyclone with a quiet center can be readily formed. Indeed, our simulation forms several GRS-scale anticylones. Furthermore, long-lived vortices in this simulation also showed a GRS-like westward drift: about 0.5 deg/day in the simulation while about 0.3 deg/day for GRS in recent times \cite{simon2018}. In the simulation, the deep dynamo sets up a slow westward drift in the low to mid latitudes. The convection plumes in the dynamo region accordingly drift westward, dragging the corresponding large vortex in the atmospheric layer. The stability of GRS, however, could not be captured in the simulation. The most stable storms in our simulation existed for about about 20 days, while the GRS has life time of 100s of years \cite{simon2018}. Possibly, Jupiter is able to sustain much longer-lived convective plumes, spawning stable GRS-like vortices, than our simulation. A broader systematic parameter study of this simulation will be useful to investigate the possibility of longer-lived large anticyclones.

Based on these two case studies we can begin piecing together a global picture of the deep-convection driven fluid dynamics occurring in Jupiter and Saturn. An atmospheric layer with dynamics governed by rapidly rotating deep convection will tend to promote either cyclonic or anticyclonic vortices/storms depending on the direction of the vorticity set by the local zonal wind shear. This phenomenon is likely active in the atmospheric layers of both Saturn and Jupiter. On top of that, if the interior structure of a planet allows the dynamo layer to directly interact with the outer hydrodynamic layer, then energetic plumes from the deep dynamo layer will exclusively excite anticyclonic vortices in the atmospheric layer, tilting the balance towards more anticyclones than cyclones. This scenario is likely active in Jupiter, explaining why it has many more anticyclones than cyclones \cite{li2004,vasavada2005}. On Saturn, however, the dynamo layer is largely decoupled from the outer hydrodynamic layer due to the likely presence of a thick stably stratified layer. Even if its dynamo layer generates energetic plumes, they get trapped by the overlying stably stratified layer, explaining the absence of a clear and consistent preference for anticyclones on Saturn \cite{vasavada2005,vasavada2006,trammell2016}. 

The vortex formation mechanisms discussed above, however, do not exclude the presence of other mechanisms that have been proposed. For example, moist convection driven shallow storms might be forming in the outermost layers of both Saturn and Jupiter. Furthermore, the idea of anticyclone formation through convective pummeling of a thin stably stratified layer from below \cite{heimpel2016} might also be active on both planets. The atmospheres of Jupiter and Saturn exhibit  extremely rich dynamics suggesting the  presence and complex interaction of various storm formation mechanisms. In this paper we offer two possibilities that appear promising. The Juno spacecraft currently in orbit around Jupiter will likely shed more light on this issue. The on-board MicroWave Radiometer (MWR) \cite{janssen2017} observes thermal radiation from levels as deep as 1000 bars. If large atmospheric vortices on Jupiter maintain some correlation with the temperature at depth, then we might be able to ascertain if some of the storms on the surface go deeper into the atmosphere.

\vspace{1.5cm}

\noindent {\bf Acknowledgements}: R.K.Y. thanks Hao Cao for interesting discussions. The work was supported by the NASA Juno project.  The computing resources were  provided by the NASA High-End Computing (HEC) Program through the NASA Advanced Supercomputing (NAS) Division at Ames Research Center and by Research Computing, Faculty of Arts \& Sciences, Harvard University. 
\vspace{0.5cm}

\vspace{0.5cm}
\noindent {\bf Author Contributions}: RKY designed and carried out the simulations. RKY, MH, and JB interpreted the results and wrote the manuscript.

\section{Methods}

\subsection{Thin shell case}

The first simulation discussed in the paper consists of a thin spherical shell bounded by a lower boundary at $0.97r_o$ and an outer boundary at $r_o$. The spherical shell rotates about a fixed axis, whose direction vector is represented by $\hat{z}$, with an angular velocity $\Omega$. The fluid dynamics in this shell is governed by thermal convection set up by a temperature contract of $\Delta T$ between the inner and the outer boundary. The fluid is assumed to be incompressible, i.e., $\nabla\cdot{\vec u}=0$, where $\vec u$ is the velocity. We use the classical Boussinesq approximation to model the fluid dynamics. Furthermore, we work with non-dimensional system of equations where the distances are scaled by the shell thickness $d=r_o - r_i$, time is scaled by $\Omega^{-1}$, and temperature is scaled by $\Delta T$. With these assumptions, the equations governing the velocity and temperature $T$ are:
\begin{gather}
 \frac{\partial\vec{u}}{\partial t}+\vec{u}\cdot\vec\nabla\vec{u}+2\hat{z}\times\vec{u}+\vec\nabla P = 
\frac{Ra\,E^2}{Pr} g(r)\,T\,\hat{r}+E\nabla^{2}\vec{u}, \label{eq:Fluid_vel}\\
\frac{\partial T}{\partial t}+{\vec{u}\cdot\vec\nabla}T  = \frac{E}{Pr}\nabla^{2}T,
\end{gather}
where $P$ is pressure. The non-dimensional gravity $g(r)$ varies as $(r_o/r)^2$ assuming that most of the planetary mass is concentrated below the inner boundary at $0.97r_o$. The relevant non-dimensional parameters are: Rayleigh number $Ra = \alpha g_o \Delta T d^3/(\kappa\nu)$, Ekman number $E=\nu/(\Omega D^2)$, and Prandtl number $Pr=\nu/\kappa$, where $\alpha$ is the thermal expansion coefficient, $g_o$ is gravity at outer boundary, $\nu$ is viscosity, and $\kappa$ is thermal diffusivity. These control parameters are set to:  $E=10^{-5}$,  $Ra = 3\times10^7$, and $Pr=0.1$. The velocity matches a stress-free condition on both boundaries. The temperature is assumed to be constant on each boundary.

The above system of equations is solved using the open source MagIC code (\href{https://magic-sph.github.io/}{{\tt https://magic-sph.github.io/}}; \cite{wicht2002}) which employs the pseudo-spectral approach. The latitudinal and longitudinal quantities are expanded using the Legendre polynomials and the radial ones are expanded using the Chebyshev polynomials. The code also uses the toroidal-poloidal decomposition to maintain strict divergence-less nature of the relevant quantities. The code uses the fast spherical harmonic library SHTns (\href{https://nschaeff.bitbucket.io/shtns/}{{\tt https://nschaeff.bitbucket.io/shtns/}}; \cite{shtns}). The equations are time-advanced using an explicit second-order Adams-Bashforth scheme for the Coriolis and the non-linear terms, and an implicit Crank-Nicolson scheme for other terms \cite{glatzmaier1984}.

We simulate the system for about 1500 rotations which was enough to show a statistically stationary behavior. To limit the computational requirements, we used four fold symmetry in the azimuthal direction \cite{al2004,heimpel2005}, which effectively makes it a quarter-spherical-wedge simulation with periodic boundary conditions on the meridional planes of the wedge. The first 930 rotations were carried out on a simulation grid of [1056, 2112, 160] where the numbers represent resolution in azimuthal, latitudinal and the radial direction, respectively. The corresponding maximum spherical harmonic degree is 1408. The remaining time period was simulated on a larger grid of [1152, 2304, 200]. Apart from sharpening the fluid dynamic features, the higher resolution did not change the simulation behavior. Despite the large grid we used, a hyperdiffusion scheme, which suppresses energies at small length scales, had to be applied to run the simulation. Following several earlier studies \cite{gastine2014zon,heimpel2005,heimpel2016}, the viscosity in our setup becomes a function of spherical harmonic degree after a certain cutoff. {\tt MagIC} multiplies the following function to the main viscous diffusion operator:
\begin{gather}
d(\ell)=1+D\left[\frac{\ell+1-\ell_{hd}}{\ell_{max}+1-\ell_{hd}} \right]^{\beta}
\end{gather}
where $D$ defines the amplitude of the function, $\ell_{max}$ is the maximum spherical harmonic degree utilized in the simulation, $\ell_{hd}$ is the degree after which the hyerdiffusion starts, and $\beta$ defines the rise of the function for degrees higher than $\ell_{hd}$. For our simulation, we use $D=10$, $\beta=6$, $\ell_{hd}=450$.

\subsection{Thick shell case}

In this simulation we use the anelastic system of equations \cite{braginsky1995,lantz1999} to model compressible, subsonic flows present in the interior of giant planets. In this approximation, the thermodynamic quantities are assumed to be a combination of a static background (tilde) and small fluctuations (prime), i.e., $x = \tilde{x}+x'$. The equations used in our simulation are in the entropy variable form with entropy contrast between the inner and the outer boundary driving the convection \cite{jones2009}. The magnetic field is scaled by $\sqrt{\tilde{\rho_o}\mu_o\lambda_i\Omega}$, where $\tilde{\rho_o}$ is density on the outer boundary, $\mu_o$ is magnetic permeability, and $\lambda_i$ is magnetic diffusivity on the inner boundary. The scales for distance and time are the same as above. With these assumptions and rescaling, the evolution for velocity is governed by:
\begin{gather}
\nabla\cdot(\tilde{\rho}{\vec u})=0 \\
\left(\dfrac{\partial \vec{u}}{\partial t}+\vec{u}\cdot\vec{\nabla}\vec{u}\right)
= -\vec{\nabla}{\frac{p'}{\tilde\rho}} 
- \dfrac{2}{E}\hat{z}\times\vec{u}
+ \dfrac{Ra}{Pr}\tilde{g} \,s'\,\hat{r} \nonumber \\
+\dfrac{1}{Pm_i\,E \,\tilde{\rho}}\left(\vec{\nabla}\times \vec{B}\right)\times \vec{B}
+ \dfrac{1}{\tilde{\rho}} \vec{\nabla}\cdot \mathsf{S}, \label{eq:vel}   
\end{gather}
where $p'$ is pressure perturbation, $\tilde{g}(r)$ is gravity (assumed to vary linearly), $s'$ is entropy perturbations, $\vec{B}$ is magnetic field, and $Pm_i$ is $\nu/\lambda_i$. The Rayleigh number $Ra$ in this formulation is given by $\alpha_o g_o T_o d^3 \Delta s(c_p \kappa \nu)^{-1}$.  The traceless rate-of-strain tensor $\mathsf{S}$ is defined by 
\begin{gather}
S_{ij}=2\tilde{\rho}\left(e_{ij}-\frac{1}{3}\delta_{ij}\vec{\nabla}\cdot\vec{u}\right) \text{with}\,\,
e_{ij}=\frac{1}{2}\left(\frac{\partial u_{i}}{\partial x_{j}}+\frac{\partial u_{j}}{\partial x_{i}}\right), \nonumber
\end{gather}
where $\delta_{ij}$ is the identity matrix. The energy conservation gives time evolution equation for entropy perturbation $s'$:
\begin{gather}
\tilde{\rho}\tilde{T}\left(\dfrac{\partial s'}{\partial t} +
\vec{u}\cdot\vec{\nabla} s' \right) =
\dfrac{1}{Pr}\vec{\nabla}\cdot\left(\tilde{\rho}\tilde{T}\vec{\nabla} s'\right) +
\dfrac{Pr\,Di}{Ra}\Phi_\nu + \nonumber \\
\dfrac{Pr\,Di\,\lambda_{norm}}{Pm_i^2\,E\,Ra}\left(\vec{\nabla}
\times\vec{B}\right)^2, \label{eq:entropy}
\end{gather}
where $\tilde{T}$ is background temperature and the viscous heating contribution is given by
\begin{gather}
\Phi_{\nu}=2\tilde{\rho}\left[ e_{ij} e_{ji} - \frac{1}{3} (\vec{\nabla}\cdot\vec{u})^2 \right].
\end{gather}
The dissipation number $Di$ is ${\alpha_o{g_o}d}/{c_p}$, where $\alpha_o , g_o$ are thermal expansivity and gravity on outer boundary and $c_p$ is specific heat at constant pressure. The evolution of the magnetic field is governed by
\begin{gather}
\dfrac{\partial \vec{B}}{\partial t} = \vec{\nabla} \times \left( \vec{u}\times\vec{B}\right)-\dfrac{1}{Pm_i}\vec{\nabla}\times\left(\lambda_{norm}\vec{\nabla}\times\vec{B}\right) \label{eq:mag}  
\end{gather}
where $\lambda_{norm}$ is local magnetic diffusivity normalized by its value at the inner boundary $r_i$. The magnetic field also follows the divergence-less condition $\vec{\nabla}\cdot \vec{B}=0$. We use the anelastic version \cite{gastine2012} of {\tt MagIC} code to simulate this system of equations.

The inner boundary is at 0.1$r_o$ and the outer is at $r_o$. The control parameters for this simulation are: $E=10^{-6}$,  $Ra = 4\times10^9$, $Pr=0.1$, $Pm_i= 1.5$. The density of the fluid changes by five density scale heights (a factor of about 150) along the radius. 

The electrical conductivity follows a step change using a hyperbolic tangent function \cite{dietrich2018}: it remains constant until about 0.85$r_o$, decreases by factors of $10^7$ in a small radius, and stays constant afterwards (See `Profile 1' is Supplementary Figure S4). The velocity matches a stress-free condition, the entropy is held constant, and the magnetic field matches a potential field on the shell boundaries. Instead of starting the simulation from small flow and magnetic field perturbations, we used a saturated dynamo simulation (without any conductivity change in its interior) from our earlier study \cite{yadav2015} and lowered its Ekman number from $10^{-5}$ to $10^{-6}$ in incremental stages. Once the $10^{-6}$  simulation showed statistical saturation in its kinetic and magnetic energy time series, we then introduced a conductivity jump in stages, reaching a decrease of $10^7$ in the outer layer.   The simulation was run for about 2000 rotations in the final form. About 50 rotations were simulated on a grid of size [2112, 1054, 400] and the rest were on a grid with [1152, 576, 400]. Here, too, increasing the resolution did not change the simulation nature substantially. This simulation also required hyperdiffusion with parameters $D=20$, $\beta=3$, $\ell_{hd}=450$ (see earlier section) for the higher resolution grid and $D=5$, $\beta=2$, $\ell_{hd}=120$ in the lower resolution grid.

%================ bibliography================
%apsrev4-2.bst 2019-01-14 (MD) hand-edited version of apsrev4-1.bst
%Control: key (0)
%Control: author (8) initials jnrlst
%Control: editor formatted (1) identically to author
%Control: production of article title (0) allowed
%Control: page (0) single
%Control: year (1) truncated
%Control: production of eprint (0) enabled
%

%=========================================
%===============EXTENDED DATA============
\renewcommand\thefigure{\arabic{figure}}
\renewcommand{\figurename}{Supplementary Figure}
\setcounter{figure}{0}

\begin{figure*}
\includegraphics[scale=0.6]{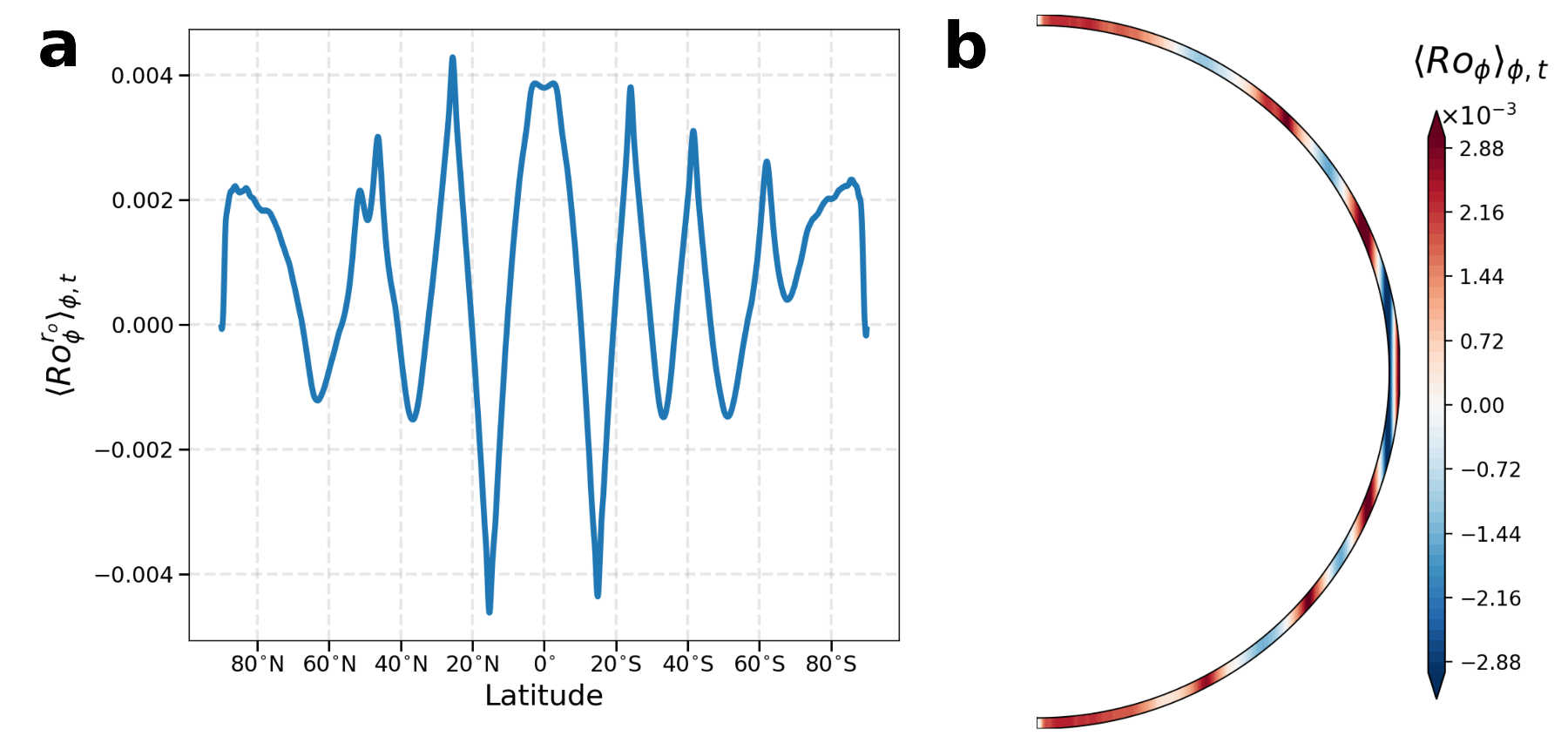}
\caption{\label{S1} The Panel {\bf a} shows the surface zonal velocity averaged in time and longitude. Panel {\bf b} shows the same as a function of latitude and radius.}
\end{figure*}

\begin{figure*}
\includegraphics[scale=0.54]{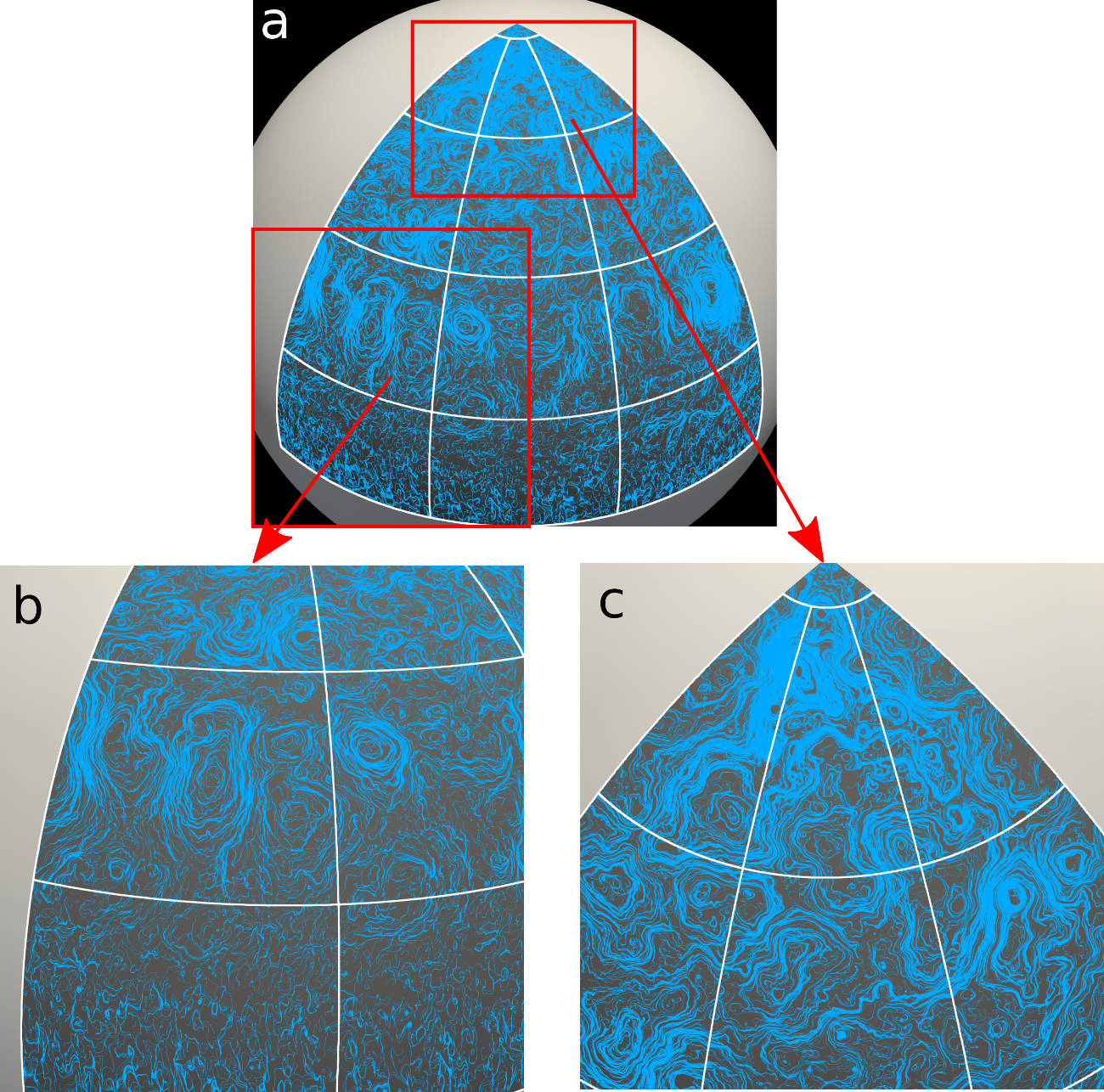}
\caption{\label{S2} The stream lines (with randomly distributed seed locations) show the flow from which zonally-averaged  azimuthal flow has been removed to highlight vortical features. Panel {\bf a} shows a global view while panel {\bf b} and {\bf c} show zoomed-in sections. The flow is shown at mid-depth radius ($0.985r_o$) of the shell. The time instant shown here is different from the instant shown in Figure 1. The low latitudes lack large scale vortical structures due to mainly jet-driven flow.}
\end{figure*}

\begin{figure*}[h!]
\includegraphics[scale=1]{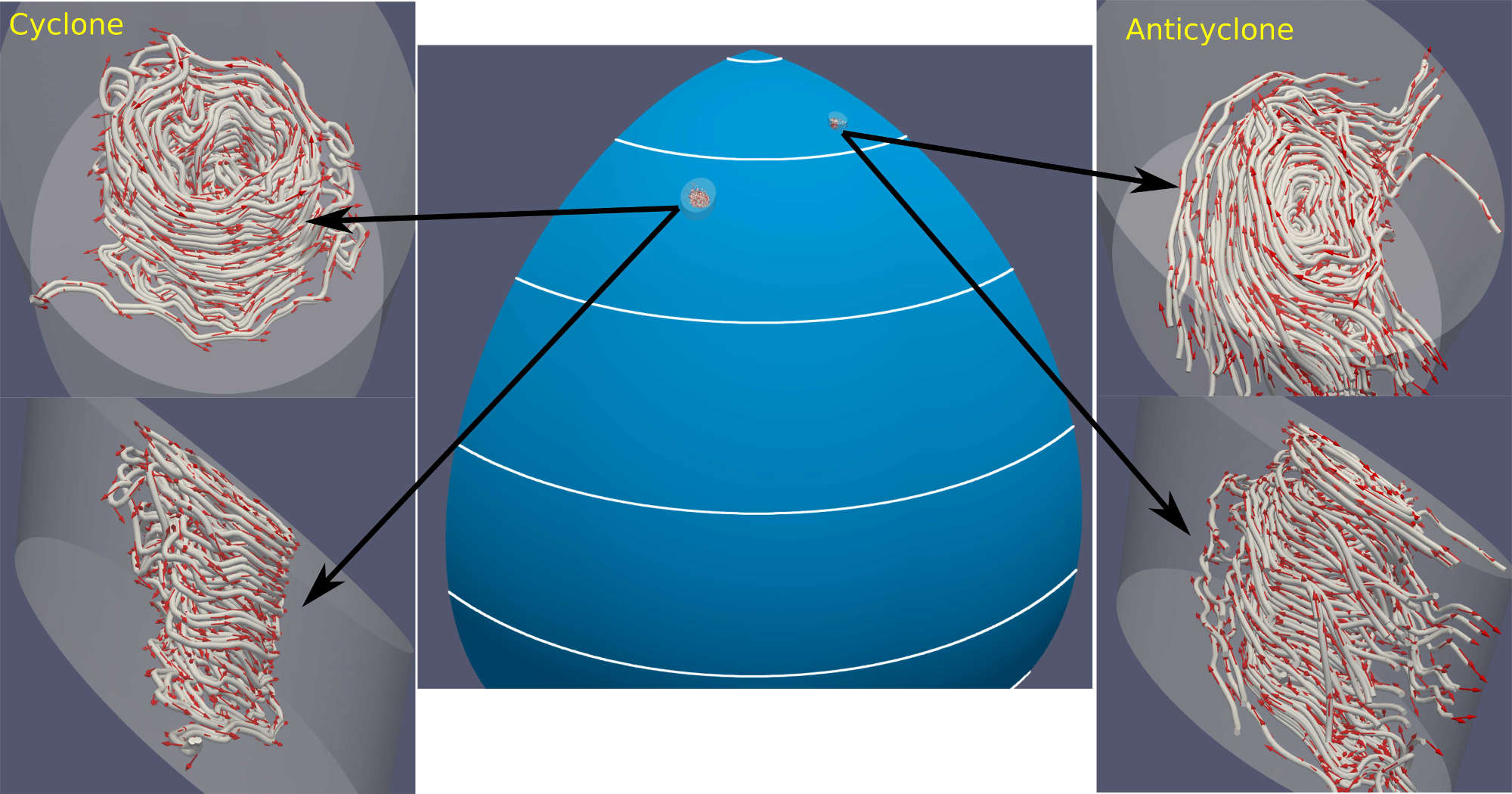}
\caption{\label{S3} The spherical wedge in the middle panel roughly spans from equator to north pole. The deep structure of a cyclone (left) and anticyclone (right) is highlighted using streamlines carrying local flow vectors. A top and side view is shown for each vortex. The grey translucent cylinders encapsulating the vortices are aligned with the planetary rotation and show the local simulation domain. }
\end{figure*}

\begin{figure*}
\includegraphics[scale=0.25]{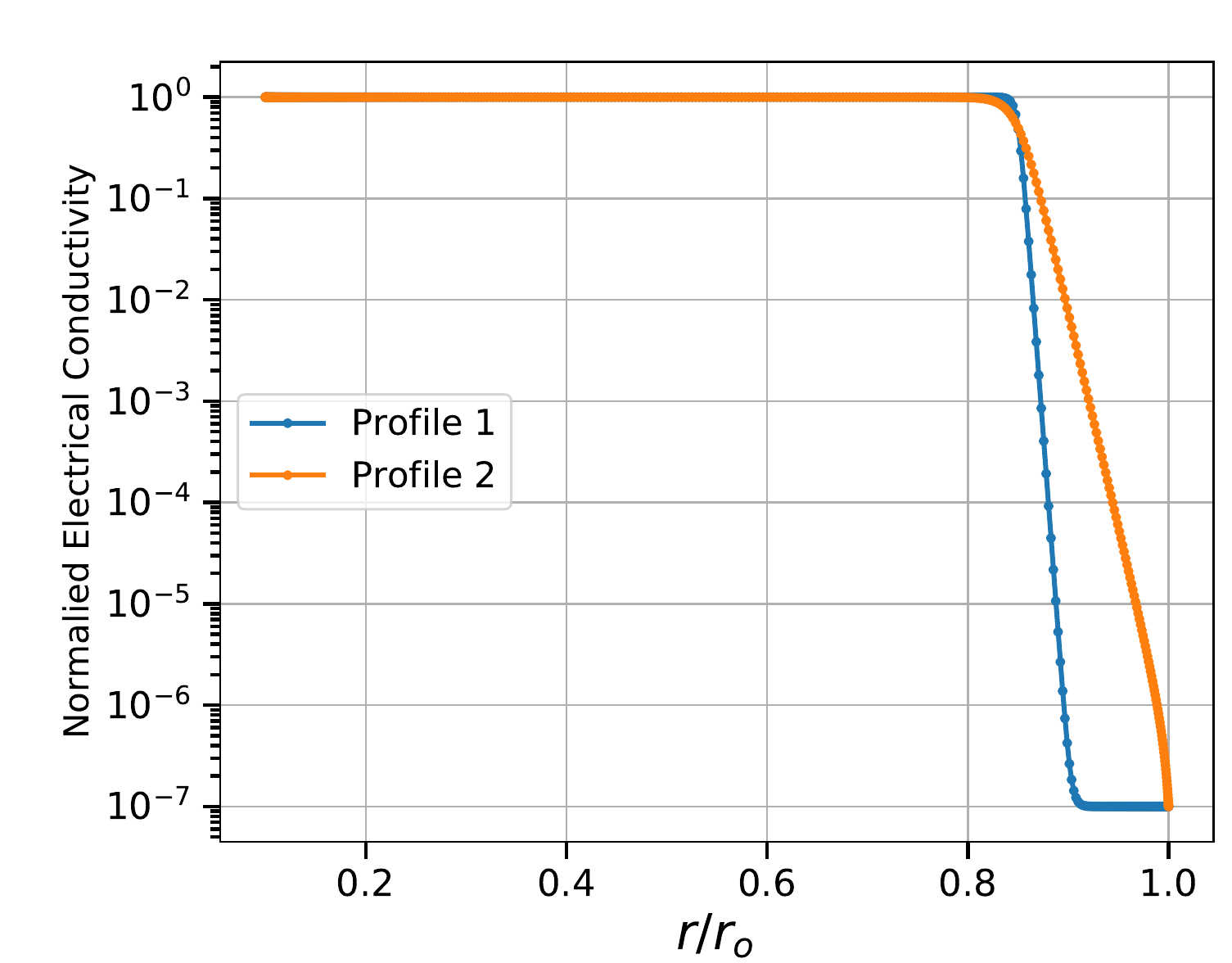}
\caption{\label{S4} The electrical conductivity (normalized by its value at the inner boundary) change as a function of radius for two dynamo simulations. The dots on the line curve denote the radial grid. The simulation with `Profile 1' shows anticyclones and is discussed in the paper.}
\end{figure*}

\begin{figure*}
\includegraphics[scale=0.5]{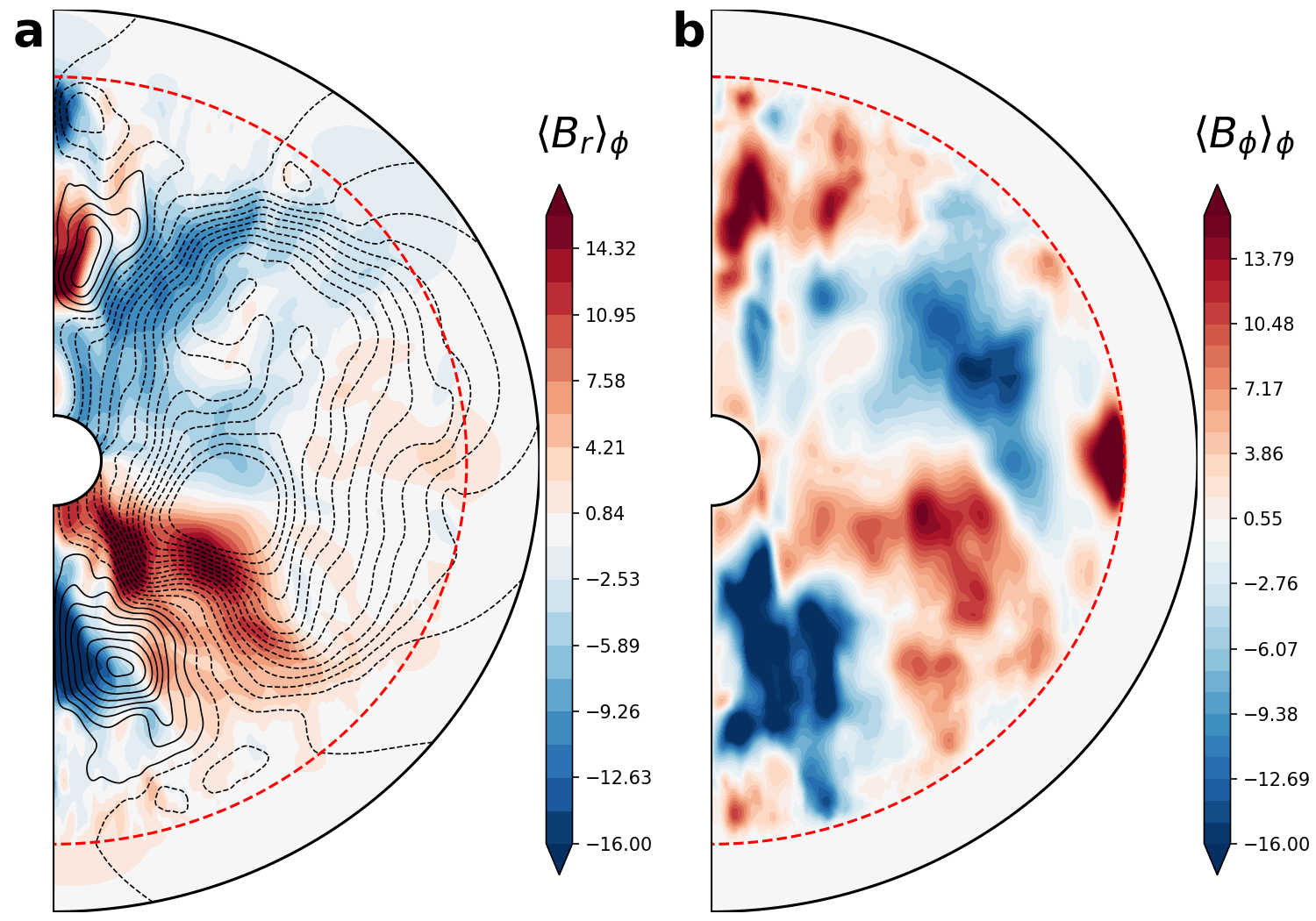}
\caption{\label{S5} Figure shows the azimuthal averages of  radial magnetic field in panel {\bf a} and of azimuthal magnetic field in panel {\bf b} at an instant of the simulation.  The poloidal magnetic field lines are also shown in panel {\bf a}. The magnetic field is in non-dimensional Elsasser number units (see the Methods section). }
\end{figure*}

\begin{figure*}
\includegraphics[scale=0.5]{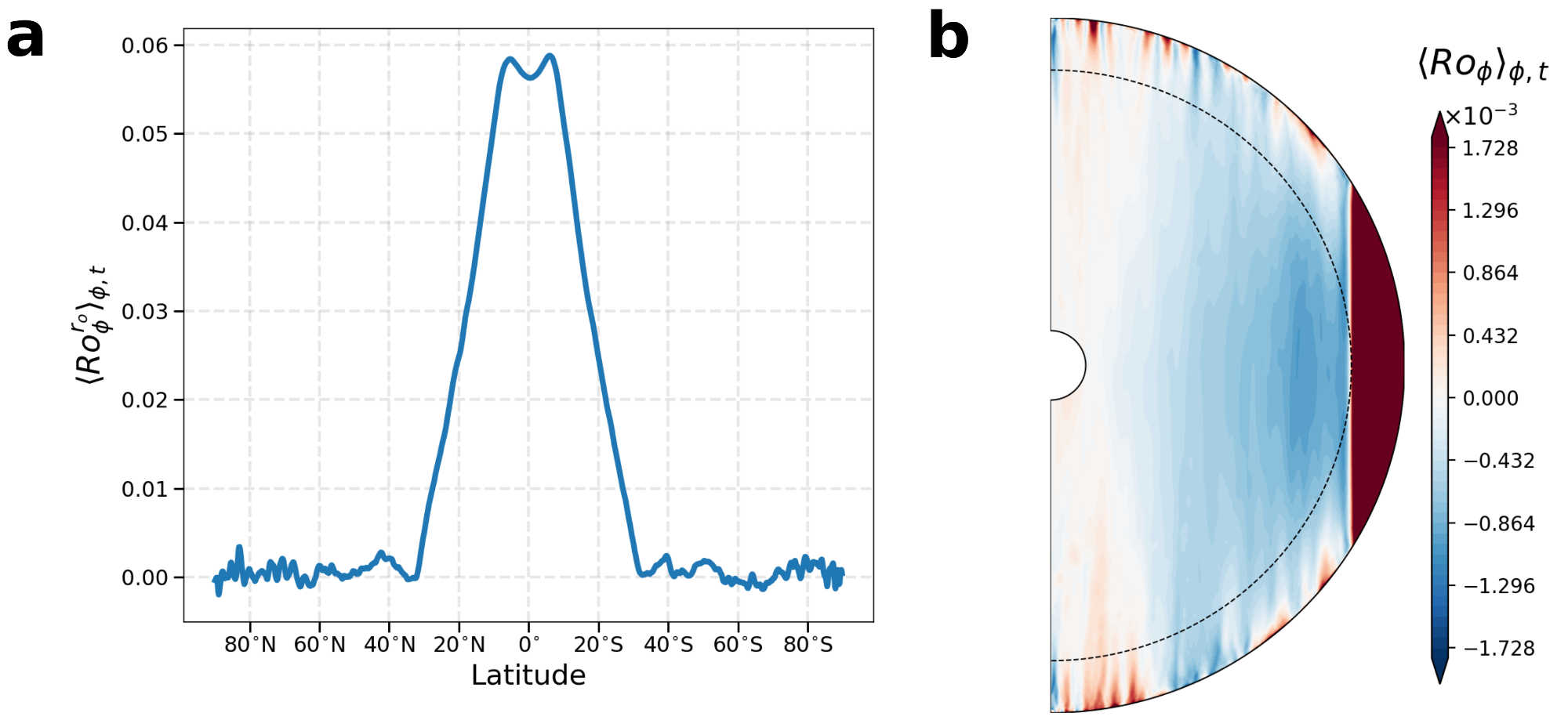}
\caption{\label{S6} Panel {\bf a} shows the surface zonal velocity averaged in time and longitude. Panel {\bf b} shows the same as a function of latitude and radius.}
\end{figure*}

\begin{figure*}
\includegraphics[scale=0.85]{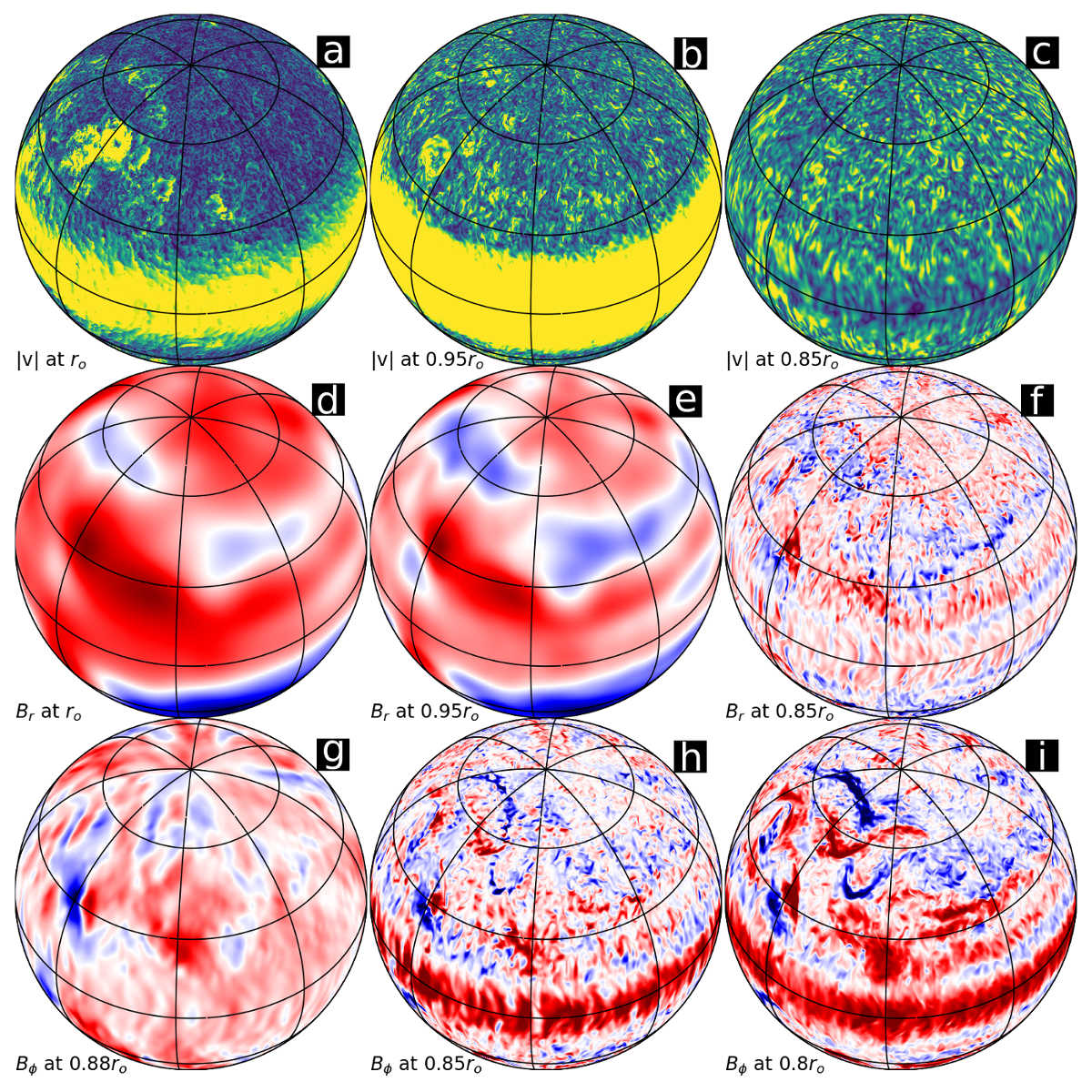}
\caption{\label{S7} Orthographic projections of the velocity magnitude and the magnetic field components at various depth. The plotted quantity is given in the lower left corner of each panel. The velocity magnitude (lighter shades representing larger values) is in terms of the Rossby number and the magnetic field (red/blue shades represent +/- values) is in the units of the Elsasser number.  The color range for different panels are: 0 to 0.045 in {\bf a}, 0 to 0.01 in {\bf b}, 0 to 0.0036 in {\bf c}, $\pm$1.5 in {\bf d}, $\pm$3  in {\bf e}, $\pm$20 in {\bf f}, $\pm$5 in {\bf g}, $\pm$15 in {\bf h}, and $\pm$30 in {\bf i}.}
\end{figure*}

\end{document}